\documentclass[pdflatex,sn-mathphys-num]{sn-jnl}

\usepackage{graphicx}%
\usepackage{multirow}%
\usepackage{amsmath,amssymb,amsfonts}%
\usepackage{amsthm}%
\usepackage[title]{appendix}%
\usepackage{xcolor}%
\usepackage{textcomp}%
\usepackage{manyfoot}%
\usepackage{booktabs}%
\usepackage{algorithm}%
\usepackage{algorithmicx}%
\usepackage{algpseudocode}%
\usepackage{listings}%
\usepackage{bookmark}%
\usepackage{tikzsymbols}
\usepackage{float}
\theoremstyle{thmstyleone}%
\theoremstyle{thmstyletwo}%

\theoremstyle{thmstylethree}%

\begin{document}

\title[Emergence of Stereotypes and Affective Polarization from Belief Network Dynamics]{Emergence of Stereotypes and Affective Polarization from Belief Network Dynamics}


\author*[1]{\fnm{Ozgur Can} \sur{Seckin}}\email{oseckin@iu.edu}\equalcont{These authors contributed equally to this work.}

\author[2]{\fnm{Rachith} \sur{Aiyappa}}
\equalcont{These authors contributed equally to this work.}

\author[3]{\fnm{Madalina} \sur{Vlasceanu}}

\author[1]{\fnm{Filippo} \sur{Menczer}}

\author[1]{\fnm{Alessandro} \sur{Flammini}}

\author*[4]{\fnm{Yong-Yeol} \sur{Ahn}}\email{yyahn@virginia.edu}

\affil[1]{\orgdiv{Observatory on Social Media}, \orgname{Indiana University}, \orgaddress{\street{1015 E 11th St}, \city{Bloomington}, \postcode{47408}, \state{IN}, \country{USA}}}

\affil[2]{\orgdiv{Luddy School of Informatics, Computing, and Engineering}, \orgname{Indiana University}, \orgaddress{\street{1015 E 11th St}, \city{Bloomington}, \postcode{47408}, \state{IN}, \country{USA}}}

\affil[3]{\orgdiv{Department of Environmental Social Sciences}, \orgname{Stanford University}, \orgaddress{\street{450 Jane Stanford Way}, \city{Stanford}, \postcode{94305}, \state{CA}, \country{USA}}}

\affil[4]{\orgdiv{School of Data Science}, \orgname{University of Virginia}, \orgaddress{\street{1919 Ivy Rd}, \city{Charlottesville}, \postcode{22903}, \state{VA}, \country{USA}}}


\abstract{
Our belief systems are shaped by social processes, such as observations and influence, and by cognitive processes, such as the drive for internal coherence. 
These processes steer how individual beliefs evolve and become connected. 
The resulting belief networks contain both causal and associative links, including spurious ones, such as stereotypes. 
Here, we develop an agent-based model of belief networks that demonstrates how two basic mechanisms---social interaction and a drive for internal coherence---can give rise to such stereotypes without any underlying reality. 
We further demonstrate how stereotypes, when coupled with shared group identity, can give rise to affective polarization, even in the absence of ideological conflicts.}

\keywords{belief systems, affective polarization, stereotypes, cognitive dissonance}



\maketitle

\section{Introduction}\label{sec1}

Beliefs shape how we perceive and engage with the world around us~\citep{ajzen1991theory, halligan2006beliefs}.
Our beliefs, along with cognitive biases, provide shortcuts and mental filters to navigate our information-rich environment and make sense of ambiguous inputs~\citep {connors2015cognitive,tversky1982judgment}.
They actively steer how we interact with others and the world. 
For instance,
we tend to talk more openly with those who share our core beliefs, while interactions with ideological outgroup members are often marked by skepticism, defensiveness, or even aggression and avoidance~\citep{iyengar2019origins,druckman2022affective}.
For example, a strong belief in the moral integrity of one's own political party often leads to giving co-partisans the benefit of the doubt during disagreements. 
By contrast, a belief in the inherent malevolence of the opposition primes us to interpret their counterarguments as deceitful or dangerous.

At the same time, beliefs are molded by what we experience both externally and internally. 
For instance, external experiences such as exposure to extreme weather events (e.g., flooding or heatwaves) reinforce existing beliefs regarding the severity of climate change~\citep{spence2011perceptions, cologna2025extreme}.
Social interactions also exert an indirect influence on our beliefs~\citep{tajfel1979integrative, asch1956studies}, as our peers reflect back their experiences that have shaped their own worldviews~\citep{bandura2001social}.
Internally, changes in some of our beliefs (e.g., political ideology) can create internal dissonance, which can then spill over to other beliefs and behaviors~\citep{mason2018uncivil, taber2006motivated, kahan2017motivated, fiorina2008political, kiley2017polarized, dellaposta2015liberals}.

Beliefs rarely exist in isolation; they interact with and are influenced by other beliefs, perceived norms, and logical constraints~\citep{converse2006nature,vlasceanu2024network,martin2002power,van1970web,turner2022belief, friedkin2016network}.
For example, parents who refuse to vaccinate their children may be more likely to oppose genetically modified foods, driven by beliefs such as ``artificial is bad and natural is good''~\citep{reich2016calling}.  
Similarly, beliefs about abortion and religiosity are strongly linked~\citep{frohwirth2018managing, pew2025abortion}. 
Although such associations are often based on direct, logical or causal links, arbitrary associations can also form and spread~\citep{goldberg2018beyond,macy2019opinion}. 
For instance, why do people who support strong gun rights tend to deny climate change? 
Why are vegetarians more likely to support more progressive income redistribution, despite no obvious link between the two attitudes~\citep{gale2007iq}? 
Or, why were liberals labeled as ``latte-drinking, sushi-eating, Volvo-driving, New York Times-reading, body-piercing, Hollywood-loving''~\citep{dellaposta2015liberals}? 
These associations can crystallize into stereotypes~\citep{kahneman1972subjective,ashmore2015conceptual}, which influence how social groups perceive one another~\citep{bordalo2016stereotypes, jussim2015stereotype, appel2021mass, czopp2015positive}, and contribute to polarization that hampers productive social discourse and fuels partisan animosity~\citep{voelkel2024megastudy, hartman2022interventions}.
Classic experiments and modeling efforts show that even minimal and arbitrary grouping cues are sufficient to produce strong in-group favoritism and outgroup bias~\citep{tajfel1971social, koster2025tabula}. This suggests that once such stereotypes attach to group identities, they can powerfully shape group judgments~\citep{rothschild2019pigeonholing, busby2021partisan} and subsequent beliefs~\citep{van2024updating}.

Explaining how arbitrary associations and stereotypes emerge is thus central to understanding the origins of strong affective polarization, which in turn may allow us to better mitigate it.
Here, we ask: how can seemingly unrelated beliefs become associated with each other, and how do these newly formed belief associations affect polarization?
Prior research showed that distinct associations can emerge from identical initial conditions, suggesting that they may be stochastic and spurious~\citep{macy2019opinion, macy2021polarization}. 
Associations can also emerge from the transmission of inferred links between behaviors (e.g., biking and hiking)~\citep{martin2014spontaneous, schultner2024transmission, uddenberg2023iterated}, which was termed ``associative diffusion''~\citep{goldberg2018beyond}.

Here, we extend a belief network model~\citep{aiyappa2024emergence,rodriguez2016collective} to demonstrate 
that stereotypes can emerge spontaneously with two simple ingredients: (i)~formation of beliefs from interactions with others and (ii)~endogenous belief updates driven by individual predisposition toward internal coherence~\citep{festinger1957theory}. 
We show that group identity can drive the emergence of shared individual preferences around initially unaligned concepts---providing a pathway to stereotypes.
Newly formed stereotypes can, in turn, intensify ingroup favoritism and outgroup derogation. Our work provides a cognitively grounded perspective on the emergence of consensus and polarization, complementing traditional models of opinion dynamics~\citep{schelling1971dynamic, steiglechner2023social, ishii2021social}.

\section{Results}\label{sec2}

\begin{figure}
    \centering
    \includegraphics[width=0.75\textwidth]{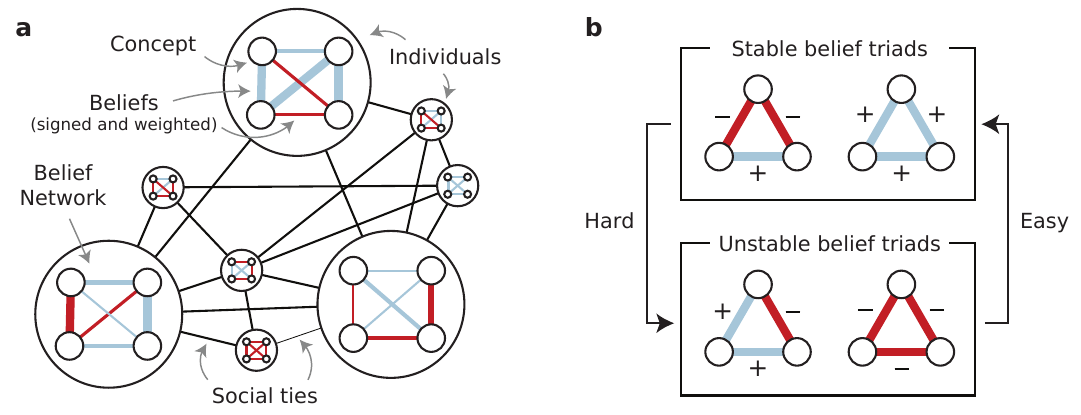}
    \caption{Configuration and dynamics of the weighted beliefs model. (a)~Individuals are represented as nodes on a social network. An individual's beliefs are also described as a network (belief network) where nodes represent concepts and weighted, signed edges between them represent beliefs, i.e., the individual's associations between two concepts. 
    (b)~Social Balance Theory is used to quantify the stability of triads within the belief network, enforcing a bias toward internal coherence.
    }
    \label{fig:summary}
\end{figure}

Consider a well-mixed social environment in which individuals can observe or communicate with each other (Fig.~\ref{fig:summary}a).
Each individual $i$'s beliefs are described as a \emph{belief network} $B_i$~\citep{aiyappa2024emergence} where nodes represent concepts.
We consider the signed and weighted edges between concepts to be ``beliefs.'' 
The sign of an edge reflects the corresponding belief's polarity, and the weight---ranging from $-1$ to $+1$---represents the strength of the belief. 
For example, the belief that Alice likes lattes would be represented as a positive edge connecting the Alice concept to the latte concept. 
Although we consider beliefs as general mental associations, they can also capture other kinds of relationships such as evidence, facts, opinions, perceived norms, or stances toward propositions~\citep{ajzen1975bayesian, schwitzgebel2011belief}. 
We adopt Social Balance Theory to operationalize the stability of triads within an individual's belief network~\citep{heider1946attitudes}, creating a natural bias toward internal coherence (Fig.~\ref{fig:summary}b)~\citep{rodriguez2016collective}.

Beliefs change through social interactions and our predisposition toward internal coherence~\citep{festinger1957theory}.
A previous model only updated beliefs communicated through social ties, based on both social and cognitive coherence mechanisms~\citep{aiyappa2024emergence}. 
However, in reality, beliefs can be indirectly and endogenously updated without explicit communication, potentially after a social interaction regarding related beliefs~\citep{goldwert2025climate}.
Previous research has highlighted the importance of incorporating belief updates that occur following social communication~\citep{goldberg2018beyond}. 
Based on this insight, we allow each social interaction to trigger an endogenous update of a related belief, driven by the predisposition toward internal coherence~\citep{houghton2020interdependent, van2019gateway, mosleh2024tendencies, nilsson2017spillover}.

\begin{figure}
    \centering \includegraphics[width=0.9\linewidth]{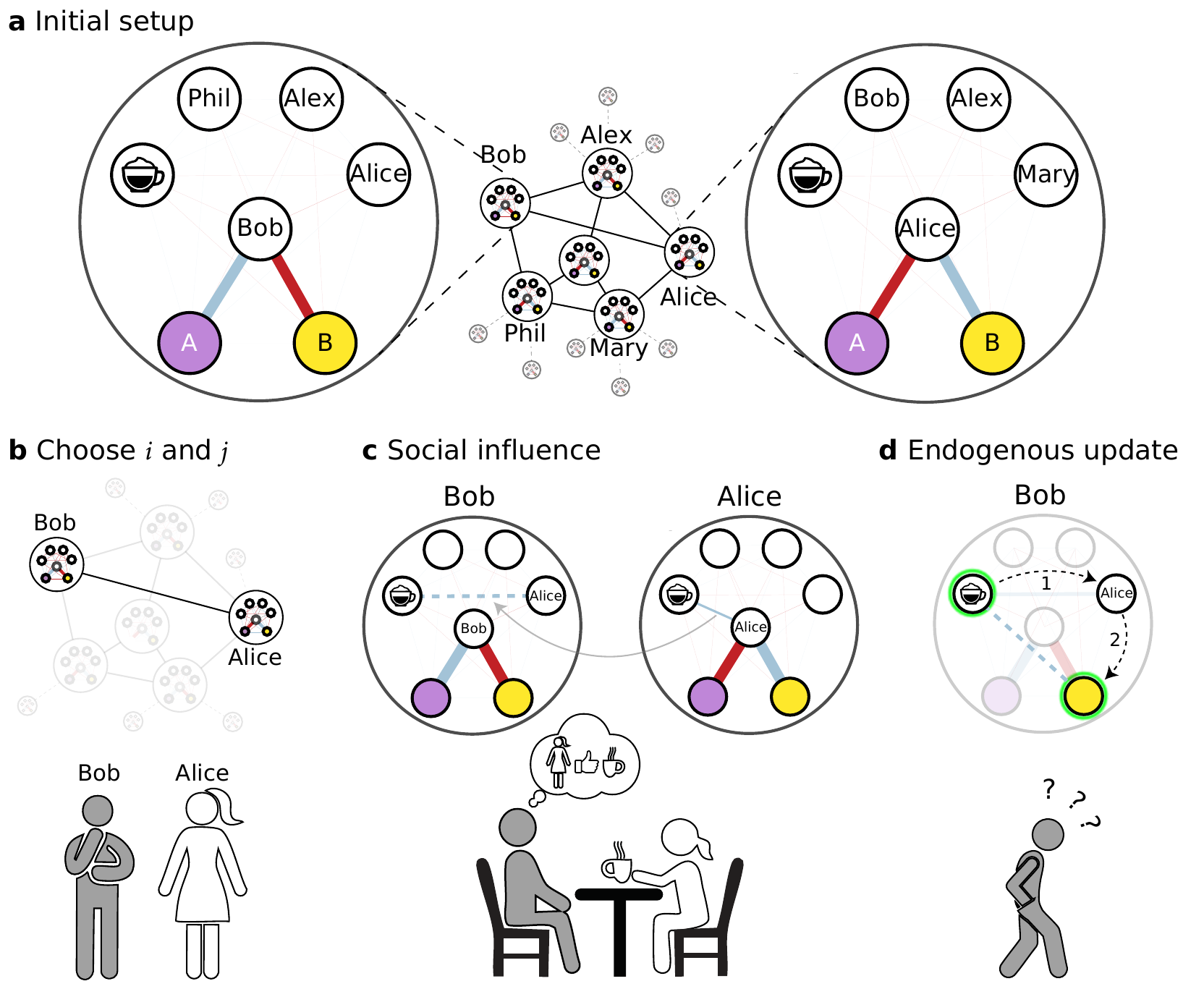}
    \caption{The setup of our model and simulation.
    The beliefs of each individual toward Group A and Group B are fixed throughout the simulation.
    (a)~The initial belief networks of two individuals---Alice and Bob---are highlighted. 
    The color of an edge indicates belief polarity, and the thickness indicates strength.
    In this example, Bob strongly supports Group A and strongly opposes Group B, while Alice strongly supports Group B and strongly opposes Group A.
    All other initial beliefs of individuals are set to be weak, sampled from $\mathcal{N}(\mu=0, \sigma=10^{-5})$.
    (b)~In each step of the simulation, an individual $i$ and its neighbor, individual $j$---Alice and Bob in this example---are selected at random.
    (c)~One belief of $j$ is randomly chosen, say $b_j(c_x, c_y)$ connecting concepts $c_x$ and $c_y$. This belief is transmitted to $i$, changing the corresponding belief in $i$'s network, $b_i(c_x, c_y)$ (dotted blue edge).
    In our example, $c_x$ is Alice and $c_y$ is latte; Bob talks to Alice, which makes him update his belief about Alice and latte.
    (d)~Triggered by this communication, the receiver $i$ then endogenously updates a related belief, which is chosen via two-step weighted random walks (dashed black arrows) starting from one of the two concepts (say $c_y$) involved in the belief $b_i$. The weighted random walks end at a set of destination nodes, from which we select one with probability proportional to the number of random walks that lead to it. In the example, Group B is selected as the destination node. Bob thus updates the belief linking Group B and latte in an effort to increase internal coherence, leading to a positive association  (dotted blue edge). 
    } \label{fig:experimental_setup}
\end{figure}

We initialize a well-mixed social network as shown in Fig.~\ref{fig:experimental_setup}a.
The nodes of an individual's belief network include a central node representing the self, nodes representing their social contacts, a prototypical concept node ``latte,'' and two nodes Group A and Group B---these could represent, for example, opposing political parties or some other group identities. 
We randomly affiliate half of the population with Group A and the others with Group B. 
As an example, Bob's belief network is illustrated in Fig.~\ref{fig:experimental_setup}a (left), where his affinity (opposition) to Group A (Group B) is indicated by a blue (red) edge, respectively. 
This belief is $+1$ ($-1$) and is fixed throughout the simulation.
All other beliefs in his network are initialized with very weak, random values close to zero (see Methods).
Individuals do not begin with any bias (other than group identity), such as favoring ingroup members, disliking outgroup members, or holding strong beliefs about shared concepts like latte. 
They also have no information about their neighbors' group affiliation or any of their other initial beliefs.

Let us now describe the dynamics of the model using the illustrations in Fig.~\ref{fig:experimental_setup}b-d.
In each time step, a receiver $i$ and a sender $j$ are selected at random. In this example, Bob and Alice are chosen as the receiver and the sender, respectively (Fig.~\ref{fig:experimental_setup}b).
Subsequently, one of $j$'s beliefs, $b_j(c_x, c_y)$, where $c_x$ and $c_y$ are concepts, is chosen at random and used to update the corresponding belief of agent $i$, $b_i(c_x, c_y)$ (see Methods).
This could be depicted as Bob learning that Alice probably likes latte, say by observing her drinking it or talking to her (Fig.~\ref{fig:experimental_setup}c).
Finally, the receiver performs an endogenous update, adjusting a belief to increase its internal coherence (see Methods). 
In the example of Fig.~\ref{fig:experimental_setup}d, the association between latte and Group A is selected to update Bob's corresponding belief in such a way that the internal coherence of his belief network increases, or equivalently its internal dissonance decreases~\citep{festinger1957theory,stieglitz2013emotions,rodriguez2016collective,aiyappa2024emergence} (see Methods). 

\subsection{Emergence of Stereotypes}

\begin{figure}[H]
    \centering
    \includegraphics[width=0.95\linewidth]{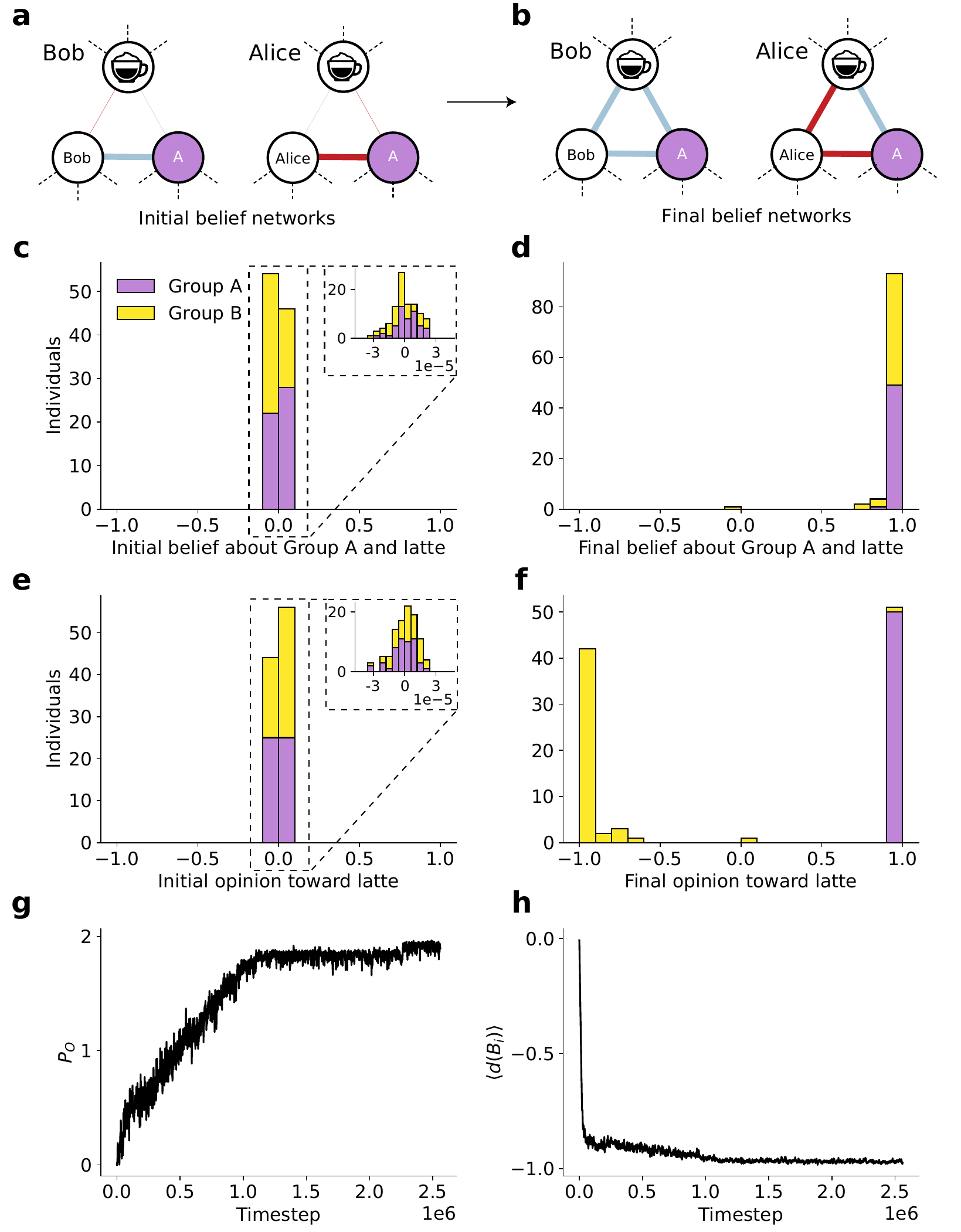}
    \caption{Even neutral concepts can be stereotyped and polarized.
    (a--b)~Stereotypes emerge from initially neutral beliefs. Alice and Bob begin with no opinions about latte, but by the end of the simulation, Bob (Group A) favors it while Alice (Group B) dislikes it, and both perceive Group A as positively associated with latte.
    (c)~Initially, there is no association between Group A and latte; beliefs are sampled from $\mathcal{N}(\mu=0, \sigma=10^{-5})$.
    (d)~By the end of the simulation, individuals across the population perceive Group A as favoring latte.
    (e)~Initially, individuals from both groups hold weak beliefs about latte, sampled from $\mathcal{N}(\mu=0, \sigma=10^{-5})$.
    (f)~By the end of the simulation, individual beliefs about latte are polarized based on group affiliation.
    (g)~Individual beliefs toward latte polarize over time ($P_O$ increases).
    (h)~The average internal dissonance ($\langle d(B_i) \rangle$) of individual belief networks decreases over time.
    }
    \label{fig:result_wout_bias}
\end{figure}

We track the emergence of stereotypes by monitoring the evolving distribution of the belief linking latte and Group A across agents.
Here, we define a stereotype as a belief that associates specific traits or preferences to a group~\citep{hilton1996stereotypes}.
Initially, individuals from both groups do not hold a belief that strongly associates latte with any group (Fig.~\ref{fig:result_wout_bias}a,c).
However, by the end of the simulation, the population exhibits a strong positive association between Group A and latte (Fig.~\ref{fig:result_wout_bias}b,d).
This stereotype arises despite the absence of any initial bias toward latte among Group A or B members (Fig.~\ref{fig:result_wout_bias}e), in line with the prior findings that such stereotypes can arise randomly in the absence of any underlying reality~\cite{macy2019opinion}.
Rather, Group A members come to prefer latte while Group B members dislike it as latte becomes positively associated with Group A (Fig.~\ref{fig:result_wout_bias}f).
This result is consistent with accounts in which stereotypes can function as self-fulfilling prophecies, where targets unconsciously align their behavior with expectations of perceivers to affirm their group identity~\cite{von1981impression, skrypnek1982self}.

Opinion polarization $P_O(c)$ around concept $c$ measures the divergence of opinions toward $c$ between Group A and B members (see Methods). 
A value of two indicates maximum polarization (i.e., the parties hold completely opposing average opinions), while a value of zero indicates no polarization (identical average beliefs). 
Fig.~\ref{fig:result_wout_bias}g tracks the evolution of opinion polarization: despite starting with unpolarized views on latte, the population ends up in a highly polarized state.  

We measure the average internal dissonance of the population, $\langle d(B_i) \rangle$ (see Methods).
Higher values indicate greater internal conflict between beliefs, while lower values suggest more internally coherent belief systems; the minimum value is $-1$ while the maximum is $+1$. 
As shown in Fig.~\ref{fig:result_wout_bias}h, the population's internal dissonance decreases over time---as polarization increases. 
This suggests that individuals converge toward internally coherent belief systems that are stable and resistant to change, thereby reinforcing and sustaining polarization at the population level.

\subsection{Emergence of Affective Polarization}

\begin{figure}
    \centering
    \includegraphics[width=.85\linewidth]{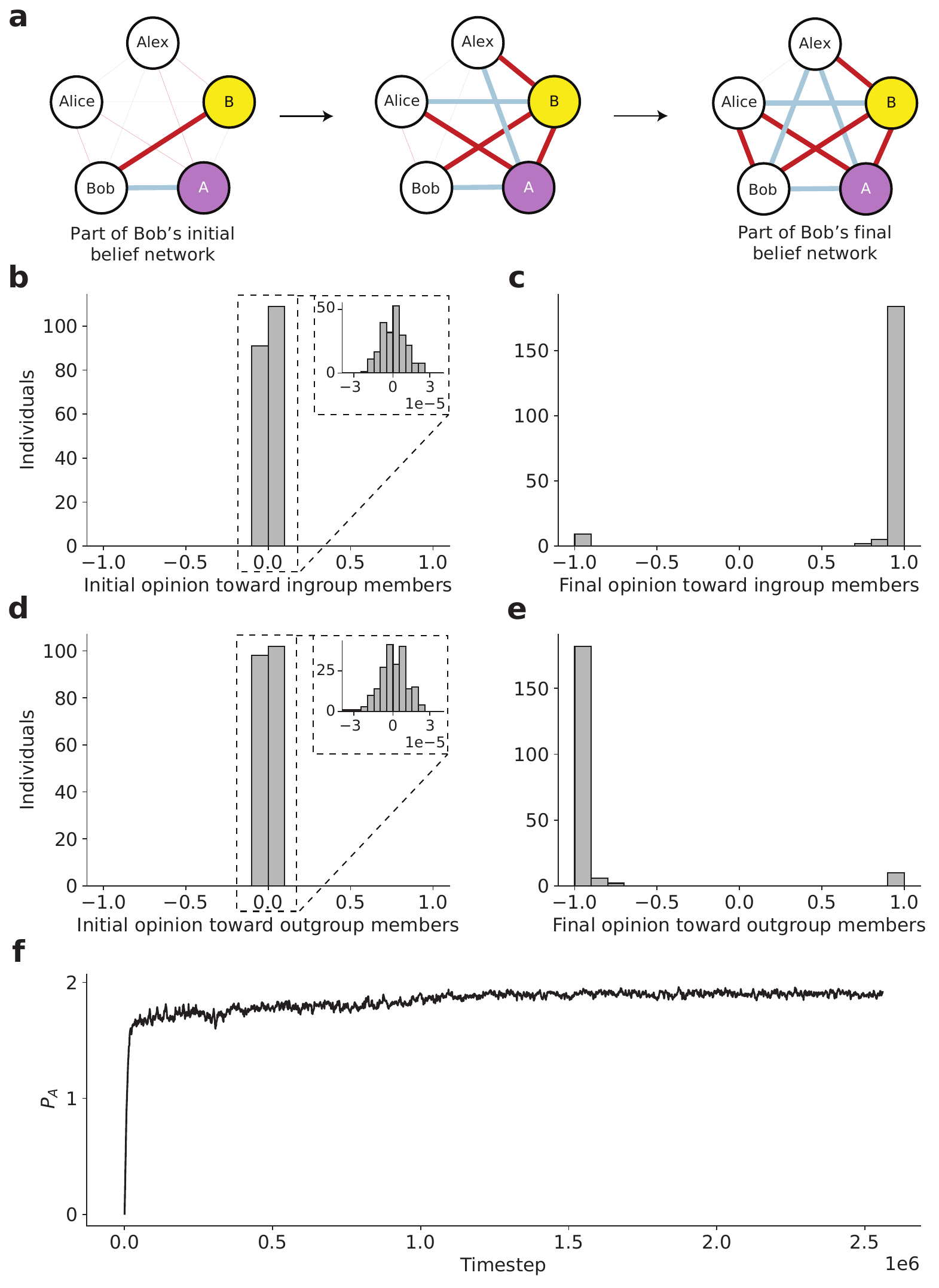}
    \caption{Affective polarization emerges from social interaction and coherence-driven belief updating.
    (a)~Ingroup favoritism and outgroup derogation develop over time.
    Initially (left), Bob holds neutral opinions about both ingroup (Alex) and outgroup (Alice) members.
    Over time (middle), Bob realizes that Alex is an ingroup member while Alice is an outgroup member. 
    Eventually (right), Bob develops strong positive opinions about ingroup members and strong negative opinions about outgroup members to maintain a coherent belief network.
    (b)~Initial and (c)~Final distributions of opinions toward ingroup members. 
    (d)~Initial and (e)~Final distributions of opinions toward outgroup members.
    Initially (b, d), individuals hold weak, unbiased beliefs about their social neighbors, with no group bias.
    By the end of the simulation (c, e), beliefs polarize: most individuals converge to $+1$ toward ingroup members and $-1$ toward outgroup members.
    (f)~Evolution of affective polarization ($P_A$), the average gap between ingroup and outgroup feelings for all individuals. It rises from 0 (neutrality) to nearly 2 (maximum polarization).
    }
\label{fig:partisan_animosity}
\end{figure}

Fig.~\ref{fig:partisan_animosity}a illustrates how Bob (a member of Group A) develops a positive belief about Alex and a negative belief about Alice by observing their beliefs and behaviors over time. 
Through this process, Bob infers that Alex supports Group A and opposes Group B, while Alice supports Group B and opposes Group A, leading Bob to feel affinity toward Alex and aversion toward Alice. 
Overall, individuals develop strongly positive beliefs (value of $+1$) about their ingroup members (Fig.~\ref{fig:partisan_animosity}c) while simultaneously forming strongly negative beliefs ($-1$) about the outgroup (Fig.~\ref{fig:partisan_animosity}e).
This occurs despite individuals starting with unbiased beliefs and no knowledge of others' group affiliations, as shown in Fig.~\ref{fig:partisan_animosity}b,d.

We also report affective polarization $P_A$, which measures the average gap between feelings toward ingroup and outgroup members within the entire population (see Methods).
Fig.~\ref{fig:partisan_animosity}f shows that affective polarization rises from $0$ to nearly $2$ over the course of the simulation.

Note that, contrary to the general pattern, Figs.~\ref{fig:partisan_animosity}c,e reveal that a few individuals develop negative beliefs about ingroup members and positive beliefs about outgroup members.
Closer inspection shows that these individuals, though a minority, \textit{falsely} infer that their outgroup (ingroup) neighbors support (oppose) their party. 
Driven by a bias toward internal coherence, they form positive (negative) beliefs about those they perceive to be politically aligned (misaligned), even when that perception is inaccurate. 
This is analogous to people overestimating the extent to which their own beliefs are shared by others, a bias known as false consensus effect~\citep{ross1977false}. When an individual experiences positive interactions with an outgroup neighbor, they may endogenously update their perception of that neighbor's political identity to align with their own. This projection allows them to resolve cognitive dissonance and maintain positive affect toward the neighbor.

In sum, our findings highlight that social influence and a predisposition toward internal coherence can not only drive opinion polarization around arbitrary concepts, but also foster ingroup favoritism and outgroup animosity---dynamics that lie at the heart of partisan conflict.

\subsection{Role of Social Influence and Belief Coherence} 

\begin{figure}
    \centering
    \includegraphics[width=1\linewidth]{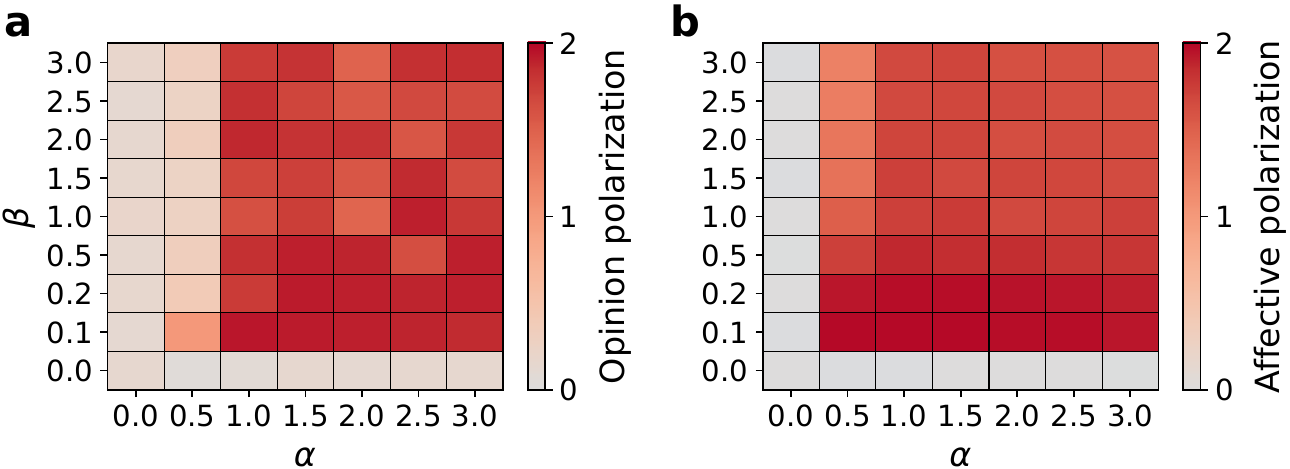}
    \caption{
    Social influence $\alpha$ and predisposition toward internal coherence $\beta$ jointly drive the emergence of group-level stereotypes and affective polarization.
    (a)~Individual beliefs about latte become polarized along group lines when $\alpha > 0.5$ and $\beta > 0$.
    (b)~Affective polarization starts emerging as $\alpha > 0$ and $\beta > 0$.
    Each cell shows average results from 10 simulation runs.}
    \label{fig:result_a_b}
\end{figure}

Fig.~\ref{fig:result_a_b} illustrates how the strength of social influence ($\alpha$) and the strength of the predisposition toward internal coherence ($\beta$) shape opinion polarization and affective polarization. 
Opinion polarization (Fig.~\ref{fig:result_a_b}a) measures the divergence of opinion toward latte between Group A and B as shown in Eq.~\ref{eq:stereotype} while affective polarization (Fig.~\ref{fig:result_a_b}b) quantifies the average gap in feelings toward ingroup and outgroup members as shown in Eq.~\ref{eq:affective}.

We report opinion polarization and affective polarization at the end of the simulations, averaged over 10 runs.
In our parameter sweep, substantial opinion and affective polarization appear only when social influence is sufficiently strong ($\alpha > 0.5$) and the drive for internal coherence is non-zero ($\beta > 0$).
A closer inspection reveals that when $0 < \alpha \leq 0.5$, individual beliefs toward latte take on values $+1$ or $-1$ regardless of group affiliation, leading to similar average beliefs within the groups and hence zero polarization.
When $\alpha > 0.5$ and $\beta > 0$, we see distributions of beliefs about latte similar to Fig.~\ref{fig:result_wout_bias}d.
Similarly, we observe distributions of beliefs about in/outgroup similar to Fig.~\ref{fig:partisan_animosity}c,d. 
Note that the predisposition toward internal coherence has a non-linear effect on polarization: for a fixed value of social influence $\alpha$, high non-zero values of $\beta$ lead to lower polarization. 
The reason is that coherence creates pressure to crystallize beliefs before the agents have time to learn associations.
In particular, false beliefs about the group memberships of neighbors lead to lower polarization. 

\section{Discussion}


While some associations made by the human mind stem from direct causal links, others are purely associative, such as the connection between certain cultural preferences and political identity.
People often search for patterns in their experiences and stitch them together to derive meaning, even when no underlying causality exists~\citep{tversky1982judgment,  murray2003narrative, shermer2011believing}.
These cognitive shortcuts can misfire, transforming spurious associations into stereotypes that falsely link groups to certain behaviors and lay the foundation for prejudice~\cite{blanco2018illusion}.

A central theoretical contribution of this work is that it explains the emergence of stereotypes, ingroup favoritism, outgroup derogation, and polarization from just two ingredients: (1) group membership, and (2) the structure and dynamics of belief networks---specifically, social interaction and coherence-seeking. Unlike theories that attribute polarization to substantive ideological disagreement, asymmetric information, or motivated reasoning~\citep{taber2006motivated, scasny2018public, leeper2014political}, our model shows that polarization can emerge in the complete absence of any underlying truth. There is no actual correlation between group membership and preferences for the neutral concept; the associations that drive division need not be normatively meaningful, politically relevant, or even accurate. This reframes polarization not as an outcome of disagreement over values, but as an emergent property of how humans organize beliefs under uncertainty~\citep{festinger1957theory, thagard2000coherence}.

This resonates with classic findings from the minimal group paradigm, where arbitrary categorization---even based on trivial criteria---is sufficient to induce ingroup favoritism and outgroup discrimination~\citep{tajfel1971social, tajfel1979integrative}. Our model provides a mechanistic account of how such dynamics arise: group labels combined with coherence-driven belief updating are enough to generate differentiated beliefs and intergroup bias.

Prevailing models typically attribute polarization to homophily or the structural separation of individuals within a network~\cite{baumann2020modeling, dandekar2013biased, axelrod1997dissemination, sasahara2021social}.
A key finding of our work is that these structural features are not necessary. Even in a well-mixed network, where individuals interact randomly regardless of group membership and with no initial knowledge of neighbors' preferences, polarization emerges purely from cognitive dynamics, without requiring echo chambers or social sorting.

The model also advances existing opinion-dynamics frameworks by shifting the unit of analysis from isolated opinions to networks of interdependent beliefs. Rather than assuming that individuals hold independent attitudes that converge or diverge through interaction~\citep{degroot1974reaching, friedkin2011social}, we model beliefs as mutually constraining structures in which changes to one association propagate across the system. This approach builds on growing evidence that attitudes are embedded in belief networks rather than stored as independent evaluations~\citep{dalege2016toward, dalege2017network}. By capturing how innocuous observations can cascade into broad patterns of stereotyping and group-based affect, the model helps explain why polarization often appears deeper and more systemic than disagreement over any single issue would predict. Importantly, the stereotypes that emerge in our model do not result from statistical learning about true population-level regularities~\citep{fiske1996stereotyping, allport1954nature}. Instead, they arise from overgeneralization from sparse social observations combined with coherence pressures that encourage beliefs to ``fit together.'' This mechanism aligns with research on illusory correlations and the formation of false group associations~\citep{hamilton1976illusory, blanco2018illusion}, but extends it by showing how such associations can stabilize and spread through social interaction even in the absence of confirmatory evidence. As a result, widely shared stereotypes can persist despite being recognized as inaccurate or anecdotal~\citep{hilton1996stereotypes}.

The model also offers a theoretical account of affective polarization that does not depend on ideological extremity or policy divergence. Empirically, partisan animosity often exceeds substantive disagreement~\citep{iyengar2012affective, finkel2020political}, a pattern that is difficult to reconcile with models centered on ideological distance alone. In our framework, negative affect toward outgroups emerges from coherence-driven belief alignment around group identity, rather than from incompatible policy positions. This suggests that affective polarization may be rooted as much in cognitive organization and social inference as in disagreement over material interests or values.

A striking real-world illustration of these dynamics may be the polarization of vaccine attitudes during the COVID-19 pandemic. Prior to the pandemic, vaccine attitudes in the United States were not strongly divided along partisan lines~\citep{baumgaertner2018partisan}. However, as the pandemic unfolded, vaccination became deeply polarized, with political identity strongly predicting vaccine uptake and attitudes~\citep{fridman2021covid, kreps2020factors}. Our model suggests that this polarization need not have arisen from genuine disagreement about vaccine efficacy or safety. Instead, it may have emerged through multiple pathways: observing prominent ingroup members taking positions on vaccination, or through indirect associative chains---such as linking lockdowns to restrictions on freedom, and subsequently connecting bodily autonomy and vaccine mandates to the same cluster of beliefs---followed by coherence-driven alignment with one's political identity. Notably, this polarization has spilled over to attitudes toward other vaccines, with declining confidence in routine childhood immunizations following the pandemic~\citep{gallup2024vaccines}---a pattern consistent with how beliefs propagate through interconnected networks in our model.


Several simplifying assumptions in the current model present opportunities for future work. First, while our use of a well-mixed social network demonstrates that polarization can emerge without community structure or homophily, real networks exhibit these features, which could further amplify polarization dynamics. Second, we consider only two opposing groups, whereas real societies feature multiple overlapping identities---political, religious, regional---that may interact in complex ways. Third, group identity is fixed throughout the simulation, though real identities can be fluid and context-dependent. Fourth, we track only a single neutral concept (latte), while real belief systems involve many interconnected concepts that may reinforce or compete with one another. Fifth, agents cannot verify beliefs against external reality, whereas some real-world beliefs are empirically testable. Sixth, beliefs are transmitted randomly between neighbors, ignoring the selective exposure and attention that characterize real information consumption. Finally, all agents exert equal influence, abstracting away from opinion leaders, media asymmetries, and differential social reach. Each of these extensions would bring the model closer to the complexity of real-world belief dynamics.

Our model treats all beliefs uniformly, without distinguishing between direct beliefs and meta beliefs (beliefs about others' beliefs). This means an individual can internalize others' perceptions of them---even inaccurate ones---a phenomenon related to gaslighting~\citep{sweet2019sociology}, though in practice individuals vary in their susceptibility to such influence~\citep{yeung2003looking}. A promising extension would incorporate theory of mind~\citep{apperly2012theory} and second-order beliefs~\citep{vlasceanu2024network}, distinguishing between how individuals perceive the world and how they believe others perceive it.

Another promising avenue for future research lies in empirically testing the model's predictions. For instance, the model predicts that exposure to a few group members exhibiting a behavior is sufficient to form group-level stereotypes, and that these stereotypes strengthen through coherence-driven reasoning even without further evidence. These predictions could be tested using controlled experiments that track belief updating over time, or through observational studies of how novel associations spread through social networks~\citep{moussaid2013social, das2014modeling, dalege2022using, batzke2025cognitive}.


Despite these simplifications, the model's mechanisms---social interaction and internal coherence---suggest that the same dynamics may underlie polarization across diverse domains, including politics, culture, science denial, and social identity more broadly~\citep{lewandowsky2017beyond}. Because the model does not depend on domain-specific assumptions, it provides a unifying theoretical framework for understanding how arbitrary symbols, behaviors, or preferences become socially and politically charged.
This generality positions the model as a foundational contribution to theories of belief formation, stereotyping, and polarization, rather than an explanation tailored to a single empirical context.

\section{Methods}

\subsection{Model}

Individuals are connected through a social network $G = (\mathcal{V}, \mathcal{E})$, where $\mathcal{V} = \{1, \ldots, N\}$ is the set of individuals and $\mathcal{E}$ is the set of social ties between them. Each individual $i \in \mathcal{V}$ has a belief network $B_i$, consisting of a set of concept nodes $\mathcal{C} = \{c_1, c_2, \ldots\}$ and a set of belief edges, where each belief $b_i(c_x, c_y) \in [-1, 1]$ represents $i$'s perceived association between a pair of concepts $c_x, c_y \in \mathcal{C}$.

Following social balance theory, for each triad $(c_x, c_y, c_z) \in \mathcal{T}$ of concepts in the belief network, the energy is given by the negative product of its three edges:
\begin{equation}\label{eq:triad_energy}
    E_i(c_x, c_y, c_z) = -b_i(c_x,c_y)\, b_i(c_x,c_z)\, b_i(c_y,c_z).
\end{equation}
A triad is balanced (negative energy) when the product of beliefs is positive, and unbalanced (positive energy) when the product is negative. The internal dissonance of individual $i$'s belief network is the average energy over all triads $\mathcal{T}$:
\begin{align}\label{eq:dissonance}
    d(B_i) &= \frac{1}{|\mathcal{T}|} \sum_{(c_x,c_y,c_z)\in\mathcal{T}} E_i(c_x, c_y, c_z) \notag \\
    &= -\frac{1}{|\mathcal{T}|} \sum_{(c_x,c_y,c_z)\in\mathcal{T}} b_i(c_x,c_y)\, b_i(c_x,c_z)\, b_i(c_y,c_z).
\end{align}
The lower the dissonance, the more coherent and stable the belief system.

The belief network evolves through pairwise social interactions and endogenous coherence adjustments. At each time step $t$, individual $i$ interacts with a randomly chosen neighbor $j \sim \mathrm{Unif}(\Gamma(i))$, where $\Gamma(i)=\{k \in \mathcal{V} : (i,k)\in\mathcal{E}\}$ is the neighbor set of $i$. At the level of individual beliefs, the update proceeds sequentially:
\begin{align}\label{eq:general}
    b_i(c_x, c_y; t+1) &= b_i(c_x, c_y; t) + f(B_i(t), B_j(t), c_x, c_y), \\
    b_i(c_m, c_n; t+1) &= b_i(c_m, c_n; t) + g(\tilde{B}_i, c_m, c_n),
\end{align}
where $(c_x, c_y)$ is a randomly chosen belief edge updated through social influence $f$, $(c_m, c_n)$ is an adjacent edge selected for internal coherence adjustment. $g$, and $\tilde{B}_i$ denotes $i$'s intermediate belief network after the social update. We now describe each component in detail.

Let us illustrate the model dynamics using Fig.~\ref{fig:experimental_setup}b-d.
At each time step $t$, two individuals Bob (individual $i$) and Alice (individual $j$) are chosen at random (Fig.~\ref{fig:experimental_setup}b).
Next, Bob learns that Alice likely likes latte, for example by observing her drinking it (Fig.~\ref{fig:experimental_setup}c).
We model this interaction formally by assuming that Alice's belief $b_\text{Alice}(\text{Alice}, \Coffeecup)$ acts as a social influence signal on Bob's belief $b_\text{Bob}(\text{Alice}, \Coffeecup)$.
This social interaction results in Bob updating his belief $b_\text{Bob}(\text{Alice}, \Coffeecup)$ about the association between Alice and latte.
The social influence term $f$ in Equation~\ref{eq:general} is given by:
\begin{equation}\label{eq:social}
    f(B_i(t), B_j(t), c_x, c_y) \sim \mathcal{N}(\alpha (b_j(c_x, c_y; t) - b_i(c_x, c_y; t)),\sigma),
\end{equation}
where $\alpha$ denotes the strength of social influence, and $\sigma$ is the standard deviation of the Gaussian noise term.
We use $\alpha=1$ and $\sigma=0.1$ unless otherwise stated. 

This update process leads Bob to adjust his belief toward Alice's belief, similar to classical opinion dynamics models in which one's opinion moves toward that of a neighbor. We also considered an alternative method in which $f(B_i(t), B_j(t), c_x, c_y) \sim \mathcal{N}(\alpha \cdot b_j(c_x, c_y; t),\sigma)$.
In this case, when two beliefs have the same polarity, they reinforce each other, implying that interactions among like-minded individuals can drive them toward more extreme views~\citep{lee2023makes, hobolt2024polarizing}. The main results remain similar under both formulations. 

The internal coherence term $g$ in Equation~\ref{eq:general} captures Bob's predisposition toward internal coherence, which causes changes in his belief system, particularly in beliefs adjacent to his new belief (Fig.~\ref{fig:experimental_setup}d).
We model this process through two-step weighted random walks on the network, where each step is taken with probability proportional to the absolute value of the corresponding edge's weight. 
Each random walk starts from one of the two concepts involved in the belief inferred by Bob upon observing Alice's social behavior (either Alice or latte, selected at random). This is the source node. The walks consider all destination concept nodes in Bob's belief network reachable in exactly two steps. 
From the resulting set of destination concepts, we select one with probability proportional to the number of weighted random walks leading to it. 
The example of Fig.~\ref{fig:experimental_setup}d shows a random walk: latte (source concept) $\overset{1}{\rightarrow}$ Alice (intermediate concept) $\overset{2}{\rightarrow}$ Group B (destination concept). Imagine that the destination selected based on all random walks is Group B. Then the association between latte and Group B would be selected to update Bob's corresponding belief. 
Bob updates the belief $b_\text{Bob}(\Coffeecup, \text{Group B})$ according to:
\begin{equation}\label{eq:coherence_update}
    g(\tilde{B}_i, c_x, c_y) \sim \mathcal{N}\left(-\beta\frac{\partial d(\tilde{B}_i)}{\partial \tilde{b}_i(c_x,c_y)}, \sigma\right),
\end{equation}
where $\beta$ denotes the strength of predisposition toward internal coherence, and we omit time $t$ to simplify the notation.
We use $\beta=1$ unless otherwise stated.
After updating a belief weight based on social influence and internal coherence, we clip belief weights to the valid range:
\begin{equation}
    b_i(c_x,c_y;t+1) \leftarrow \max\{-1, \min[1,\, b_i(c_x,c_y;t+1)]\}.
\end{equation}
The derivative of the internal dissonance (Equation~\ref{eq:dissonance}) with respect to focal belief $b_i(c_x,c_y)$ is given by
\begin{equation}
\frac{\partial d(B_i)}{\partial b_i(c_x,c_y)}
=
-\frac{1}{|\mathcal{T}|}
\sum_{c_z: (c_x,c_y,c_z) \in \mathcal{T}_{(c_x,c_y)}}
b_i(c_x,c_z)\, b_i(c_y,c_z)
\label{eq:derivative}
\end{equation}
where the sum is over all triads $\mathcal{T}_{(c_x,c_y)}$ that include edge $(c_x,c_y)$.
To get an intuition for how Equations~\ref{eq:coherence_update}, \ref{eq:dissonance}, and \ref{eq:derivative} work together in the model, let us consider the case in which the focal belief $b_i(c_x,c_y)$ is \textit{positive} and all the triads that include it ($\mathcal{T}_{(c_x,c_y)}$) are stable.
In this case, the terms of the sum in Equation~\ref{eq:derivative} are positive (the other edges in each triad are either both positive or both negative).
The derivative $\frac{\partial d(B_i)}{\partial b_i(c_x,c_y)}$ is therefore negative, and according to Equation~\ref{eq:coherence_update}, $b_i(c_x,c_y)$ will increase. This will decrease his internal dissonance $d(B_i)$, per Equation~\ref{eq:dissonance}, therefore increasing the overall coherence of the belief system $B_i$.
If the triads in $\mathcal{T}_{(c_x,c_y)}$ are unstable, the derivative $\frac{\partial d(B_i)}{\partial b_i(c_x,c_y)}$ is positive (Equation~\ref{eq:derivative}) and $b_i(c_x,c_y)$ will decrease (Equation~\ref{eq:coherence_update}), again decreasing the dissonance $d(B_i)$ (Equation~\ref{eq:dissonance}) and increasing the overall belief coherence.
Similarly, stable triples will tend to decrease a \textit{negative} focal belief while unstable triples will tend to increase it.

\subsection{Population-Level Metrics}

\subsubsection{Opinion Polarization}

Let $\mathcal{A}$ and $\mathcal{B}$ denote the sets of Group A and Group B members, respectively, and let $b_i(i, \Coffeecup)$ denote individual $i$'s opinion about latte (i.e., the belief about the association between self and latte). We calculate opinion polarization as the absolute difference between the mean opinions of the two groups:
\begin{equation}\label{eq:stereotype}
    P_O(\Coffeecup) = \left| \frac{1}{|\mathcal{A}|} \sum_{i \in \mathcal{A}} b_i(i, \Coffeecup) - \frac{1}{|\mathcal{B}|} \sum_{j \in \mathcal{B}} b_{j}(j, \Coffeecup) \right|
\end{equation}

\subsubsection{Affective Polarization}

Let $b_i (i,j)$ denote individual $i$'s valence toward individual $j$.
Let $G_\text{in}(i)$ and $G_\text{out}(i)$ denote the sets of $i$'s ingroup and outgroup members, respectively, excluding $i$ itself.
For each individual $i$, we define the mean valence toward ingroup and outgroup members as:
\begin{equation}
    \langle b_i \rangle_\text{in}  = \frac{1}{|G_\text{in}(i)|} \sum_{j \in G_\text{in}(i)} b_i(i,j),
    \qquad
    \langle b_i \rangle_\text{out} = \frac{1}{|G_\text{out}(i)|} \sum_{j \in G_\text{out}(i)} b_i(i,j),
\end{equation}
Then, affective polarization for the population in a simulation run is:
\begin{equation}\label{eq:affective}
P_A = \frac{1}{N}\sum_{i=1}^N \left( \langle b_i \rangle_\text{in} - \langle b_i \rangle_\text{out} \right)
\end{equation}
where $N = |\mathcal{A}|+|\mathcal{B}|$.

\subsubsection{Internal Coherence}

The internal dissonance $d(B_i)$ of agent $i$'s belief network $B_i$ is measured using social balance theory (Eq.~\ref{eq:dissonance}).
We average it across all individuals to obtain the population-level dissonance, $\langle d(B_i) \rangle = \frac{1}{N} \sum_{i} d(B_i)$.
Lower the value of dissonance, higher is the internal coherence. 

\subsection{Simulation}

We create a social network with $N=100$ individuals and $M=200$ edges randomly distributed among them.
Each individual exhibits an internal belief network.
In each agent's internal belief network, the weight of the link from the self-node to the supported group (opposed group) is fixed at $+1$ ($-1$).
We initialize weights between all other nodes in the belief system by sampling from a normal distribution $\mathcal{N}(\mu=0, \sigma=10^{-5})$. This initialization yields fully connected belief networks.

We run the simulations for 2.5 million time steps to make sure that the population-level metrics (i.e., $P_O$, $P_A$ and $d(B_i)$) have approximately reached a stable steady state.

\bmhead{Code Availability Statement}
We developed custom code using Julia for the model, and Python for data analysis.
The replication code is available via GitHub.\footnote{\url{https://github.com/rachithaiyappa/emerging_beliefs}}

\backmatter

\bmhead{Acknowledgements}

O.C.S., R.A., M.V., F.M., A.F., and Y.-Y.A. were supported in part by the Air Force Office of Scientific Research under award no. FA9550-25-1-0087. 
R.A. was also supported in part by the Institute for Humane Studies (under grant number IHS017931)
This work was also supported in part by the Knight Foundation.
This work used the IU JetStream 2 computational infrastructure through allocation CIS240118 from the Advanced Cyberinfrastructure Coordination Ecosystem: Services \& Support (ACCESS) program, which is supported by National Science Foundation grants \#2138259, \#2138286, \#2138307, \#2137603, and \#2138296 \cite{hancock2021jetstream2, boerner2023access}.

\bibliography{sn-bibliography}

\end{document}